\documentclass[letterpaper]{JHEP3}
\usepackage{array}

\usepackage{nicefrac}
\usepackage{amsmath}

\usepackage{amssymb,amsfonts,mathrsfs}
\usepackage{graphicx}
\usepackage{epsfig}

\usepackage{amsfonts}

\newcommand{\Tr}{\mbox{Tr}}

\def\h{\eta}

\def\IC{\relax\hbox{$\inbar\kern-.3em{\rm C}$}}

\def\IC{{\bf C}}

\def\bea{\begin{eqnarray}}
\def\eea{\end{eqnarray}}
\def\be{\begin{equation}}
\def\ee{\end{equation}}
\def\ba{\begin{align}}
\def\ea{\end{align}}
\def\bse{\begin{subequations}}
\def\ese{\end{subequations}}
\def\1F1{{}_1\!F_1}
\def\2F0{{}_2\!F_0}

\def\h3{$\textrm{H}_3^+$}

\def\IC{{\mathbb C}}

\def\Tr{{\rm Tr}}

\def\lbldef#1#2{\expandafter\gdef\csname #1\endcsname {#2}}

\def\href#1#2{#2}

\newcommand{\beq}{\begin{equation}}
\newcommand{\eeq}{\end{equation}}
\newcommand{\ber}{\begin{eqnarray}}
\newcommand{\eer}{\end{eqnarray}}

\def\be{\begin{eqnarray}}
\def\ee{\end{eqnarray}}

\def\({\left(}
\def\){\right)}
\def\[{\left[}
\def\]{\right]}
\def\<{\langle}
\def\>{\rangle}

\def\R{\mathbb R}
\def\S{\mathbb S}
\def\T{\mathbb T}

\title{On the reduction of $4d$ ${\cal N}=1$ theories on $\S^2$}
\author{Abhijit Gadde,$^\alpha$ Shlomo S. Razamat,$^{\beta,\gamma}$ and Brian Willett$^\alpha$
\\
\\
$^\alpha$\it  IAS, Princeton, NJ 08540, USA\\
$^\beta$\it NHETC, Rutgers, Piscataway, NJ 08854, USA\\
$^\gamma$\it Department of Physics, Technion, Haifa 32000, Israel\\
}

\bigskip

\abstract{

We discuss reductions of general ${\cal N}=1$ four dimensional gauge theories on $\S^2$.
The effective two dimensional theory one obtains depends on the details of the coupling of the theory to background fields, which can be translated to a choice of R-symmetry. We argue that, for special choices of R-symmetry, the resulting two dimensional theory has a natural interpretation as an ${\cal N}=(0,2)$ gauge theory.  As an application of our general observations, we discuss reductions of ${\cal N}=1$ and ${\cal N}=2$ dualities and argue that they imply certain two dimensional dualities.

\

}

\begin{document}

\section{Introduction}

Supersymmetric quantum field theories in various numbers of spacetime dimensions have been the subject of intense study in the last few decades, and they exhibit many interesting properties.  One striking example is the phenomeon of duality, where two naively quite different QFTs are actually equivalent, or flow to the same CFT at low energies.  In attempting to organize the web of known dualities, one observes many apparent relations between dualities in different numbers of dimensions.  It is natural to try to explain such relations by analyzing compactifications of the higher dimensional theories, but there are many subtleties in this analysis, as discussed for example in \cite{Aganagic:2001uw,Aharony:2013dha,Aharony:2013kma}.  In the present paper we attempt to understand the compactification of four dimensional ${\cal N}=1$ theories on $\S^2$ to obtain two dimensional theories with (at least) ${\cal N}=(0,2)$ supersymmetry.  In particular, given a four dimensional duality, we can find evidence for the existence of a corresponding two dimensional duality.

Given two four dimensional theories related by an IR duality, we can compactify some of the spacetime dimensions on a compact manifold ${\cal M}$.  At scales below the size of ${\cal M}$, the two theories should look as effective lower dimensional theories.  If, moreover, the compactification scale associated to ${\cal M}$ is sufficiently small ({\it i.e.}, ${\cal M}$ is sufficiently large), each of the theories can be approximated by their IR fixed point at the scale of the compactification, and thus the lower dimensional theories should be equivalent.  In some situations these effective theories may have useful lower dimensional UV completions, which will then be IR dual.  Such a completion might or might not coincide with the naive dimensional reduction of the four dimensional theory, {\it i.e.}, the theory we get by taking ${\cal M}$ infinitely small. In other words, the limits of going to infrared and sending the size of the compact manifold to zero might or might not commute.\footnote{Note that even in the case of a conformal duality, taking the size of ${\cal M}$ to zero in the classical action need not produce the same theory as flowing to low energies on the compactified geometry.}  Our main result in this paper is derivation of a necessary condition for the two limits to commute in the case of $4d$ ${\cal N}=1$ theories compactified on $\S^2$. We then will discuss several reductions with this condition satisfied and bring arguments in favor of certain two dimensional dualities, some of which are well known and some of which are new. 

An important set of tools in understanding supersymmetric QFTs are the supersymmetric partition functions on compact manifolds, which can often be computed exactly by localization.  These can also be used to gain some insight into compactification, since in some cases partition functions defined in higher dimensions approach those in lower dimenisons as certain limits of the geometry are taken.  For example, one can take a limit of the $\S^3 \times \S^1$ index of a four dimensional theory as the radius of the $\S^1$ is taken to zero, and in this way one obtains the $\S^3$ partition function of its compactification \cite{Imamura:2011uw,Gadde:2011ia,Dolan:2011rp}.  The main tool of our analysis here will be the $\S^2 \times \T^2$ partition function computed recently in \cite{Benini:2015noa,Honda:2015yha}, and its relation to the elliptic genus of the two dimensional theory we obtain by compactification.  We will use these to derive the condition under which the limits above commute.

Let us state the main result here.  If we take a $4d$ ${\cal N}=1$ theory with a choice of $U(1)_R$ symmetry such that all chiral multiplets have {\it non-negative integer} R-charges, then the effective $2d$ theory we obtain after compactifying on $\S^2$ with a certain twist is described by an ${\cal N}=(0,2)$ gauge theory with the same gauge group, and with matter content determined by eq. (\ref{redrul}).

The outline of this paper is as follows.  In section $2$ we describe the compactification of four dimensional ${\cal N}=1$ theories on $\S^2$.  To study this we will use in a crucial way the elliptic genus and $\S^2 \times \T^2$ partition function, which we review.  We derive a rule for determining a two dimensional ${\cal N}=(0,2)$ theory describing the compactification of a given four dimensional ${\cal N}=1$ theory.  In the subsequent sections we apply this observation to various four dimensional dualities to find evidence for two dimensional dualities.  In section $3$ we consider Seiberg duality, and show that it implies the $U(N)$ triality of \cite{Gadde:2013lxa}, as well as a new $SU(N)$ duality.  In section $4$ we consider theories with ${\cal N}=2$ supersymmetry, and find they can reduce to two dimensional theories with additional supersymmetry, {\it i.e.}, ${\cal N}=(0,4)$ or ${\cal N}=(2,2)$.  

\

\section{Reduction of four dimensional theories to two dimensions}

Let us consider a four dimensional ${\cal N}=1$ theory placed on the curved background $\S^2 \times \R^2$.  The metric is,

\be ds^2 = R^2( d\theta^2 + \sin^2 \theta d \phi^2)+ dx^2 + dy^2\,. \ee
At low energies compared to $R^{-1}$, only low energy excitations on the $\R^2$ will survive, and these will be described by an effective two dimensional theory.  We will argue that in certain cases, this theory will be supersymmetric, and we can determine its matter content.  More precisely, we will determine the matter content of a $2d$ UV description which flows to the same fixed point.

In order to preserve some supersymmetry, as in \cite{Closset:2013sxa,Benini:2015noa,Honda:2015yha} we consider theories which possess a $U(1)$ R-symmetry, and turn on a background R-symmetry gauge field,

\be \label{rsm} A = \frac{1}{2} \cos \theta d \phi\,, \ee
which has unit magnetic flux through the $\S^2$.  Then in the references above it was shown that this background preserves two supercharges, which transform as scalars under diffeomorphisms of the $\S^2$, and have the same chirality on the $\mathbb{R}^2$.  Thus the effective theory at low energies is a two dimensional theory with ${\cal N}=(0,2)$ supersymmetry.\footnote{In \cite{Kutasov:2013ffl} a similar setup was considered, where one compactifies a $4d$ ${\cal N}=1$ theory on $\T^2$ with flux turned on for background flavor symmetries (rather than the $R$-symmetry), and one obtains an effective $2d$ ${\cal N}=(0,2)$ description.  This will lead to different $2d$ theories than the ones we obtain here.}

Because of the R-symmetry flux, the R-symmetry we choose must assign integer charges to all of the basic fields in the Lagrangian, so that they take values in well-defined bundles over $\S^2$.  In practice this means we must take a combination of the superconformal R-symmetry with various $U(1)$ flavor symmetries, including those in the maximal torus of non-abelian flavor symmetries, which will typically break some of the flavor symmetry.  As we will see below, this choice of R-symmery will play a large role in determining the matter content of the resulting two dimensional theory.

Consider a single free chiral multiplet of R-charge $r$.  Then this chiral feels a magnetic flux $r$ on the $\S^2$, and the KK modes on the $\S^2$ of its component fields can be expanded in certain monopole spherical harmonics.  Taking only the zero mode components, it was shown in \cite{Closset:2013sxa} that the modes along the $\R^2$ organize themselves into $2d$ fields by the following rule,

$$ 4d\; {\cal N}=1 \; \mbox{chiral multiplet of R-charge} \; r \rightarrow $$
$$ 1-r \;\; \times \;\; {\cal N}=(0,2)\; \mbox{chiral multiplets}, \;\;\;r<1 $$
\be \label{redrul} r-1 \;\;\times\;\; {\cal N}=(0,2) \; \mbox{fermi multiplets}, \;\;\; r>1 \ee
and there is no contribution if $r=1$.

Next consider a gauge theory.   If we consider a limit where the $\S^2$ is very small, then naively one expects configurations where the gauge multiplet has any non-trivial dependence on the $\S^2$ to be strongly suppressed in the path integral by the Yang-Mills term:\footnote{Here $D_\mu$ is a derivative covariant with respect to diffeomorphism, gauge, and $R$-symmetry transformations, with the connection (\ref{rsm}) above.}

\be \label{YM} S_{YM} = \int_{\S^2 \times \R^2} d^4 x \mbox{Tr}( \frac{1}{4} F_{\mu \nu}F^{\mu \nu}  + \frac{i}{2} \lambda^\dagger \gamma^\mu D_\mu \lambda  + \frac{1}{2} D^2) \ee
The modes of the vector multiplet along the $\R^2$ can be seen by a similar analysis as above to give rise to an ${\cal N}=(0,2)$ vector multiplet in two dimensions.  Thus in this limit, one expects to find a $2d$ gauge theory with the same gauge group as the $4d$ theory, and with the matter content determined from the $4d$ matter content by applying (\ref{redrul}) to each chiral multiplet~\cite{Yujinote}.

However this argument turns out to be a bit fast.  As argued in \cite{Benini:2015noa} in the context of the $\S^2 \times \T^2$ partition function, there may be BPS configurations with a constant flux for the gauge field through the $\S^2$ which contribute to the path integral, despite naively being suppressed by the Yang-Mills term.  In their computation, this is achieved by shifting the contour of integration of the field $D$ in the ${\cal N}=1$ vector multiplet to complex values such that it cancels against the contribution from $F_{\mu \nu}$ in the classical contribution (\ref{YM}) (see also \cite{Closset:2015rna}).  These configurations comprise saddle points of the path integral which, though they lie off the contour of integration, may still contribute to the result, and a careful analysis shows that they indeed sometimes do.  Thus we must consider more carefully the contribution of such flux sectors when we take the two dimensional limit.

We will take the computation of the $\S^2 \times \T^2$ partition function as a crucial guide in determining when these flux sectors need to be taken into account in the reduction.  As above, one performs a topological twist on the $\S^2$, and as a result the partition function is independent of the size of the $\S^2$.  Thus it can be reinterpreted as the $\T^2$ partition function, or elliptic genus, of the $2d$ theory we obtain by reducing on the $\S^2$.  As we will see below, the zero flux sector of the $\S^2 \times \T^2$ index looks indentical to the elliptic genus of the $2d$ theory we obtain by the naive procedure described above, but in general there may be other flux sectors which contribute.  However, in some cases one finds that the contribution of all non-zero flux sectors vanishes, and only the zero flux contribution remains.  In such cases, we propose that the naive reduction gives the correct description of the reduced theory.  Indeed, this conclusion is consistent (by construction) with the interpretation above that the $\S^2 \times \T^2$ index computes the elliptic genus of the reduced theory.  The matching of elliptic genera of $2d$ theories is a strong indication of their equivalence.  For example, it implies that all mixed 't Hooft anomalies between flavor symmetries, and between the R-symmetry and flavor symmetries, agree.\footnote{However, it should be stressed that the elliptic genus is insensitive to the $D$-terms in the action, and so, {\it e.g.}, does not detect the metric at infinity on the moduli space, which one would expect to match for dual theories.  We will not address the role of $D$-terms in the dualities we consider below, but see \cite{3d2d} for a related discussion.}  Below we will derive the necessary condition that the zero flux sector is the only one which contributes.

In cases where non-zero fluxes do contribute, one no longer finds an expression for the elliptic genus which look like that of a single $2d$ ${\cal N}=(0,2)$ theory, but rather it looks like a direct sum of such theories, with one for each BPS gauge flux configuration.  It is an important problem to understand the two dimensional origin of the terms coming from the higher flux sectors, and  though we leave the precise resolution of the problem to future work, let us briefly present a speculation. The zero flux sector can be understood as the sector of zero-modes coming from the Kaluza-Klein reduction on $\S^2$ with R symmetry flux. The other flux sectors in two dimensions could come from defect operators that wrap the $\S^2$ and support a transverse gauge flux. If quantizing their moduli space produces the same spectrum that one can read off from the elliptic genus then it will provide a good support to this speculation. In this picture, the local operators in $2d$ associated to a given flux sector are closed under operator product expansion because their parent surface operators wrap homologous cycles and hence support the same gauge flux. This could explain why the $\S^2 \times \T^2$ partition function is  the elliptic genus of  a direct sum of two dimensional theories.  Below we will focus on the less exotic cases where only a single flux sector contributes.\footnote{See \cite{3d2d} for further discussion on the appearance of direct sums of theories when reducing to two dimensions.  For appearance of direct sums of theories in different contexts see for example~\cite{Witten:1995zh,Sharpe:2014tca}.}

Let us now review the partition functions which we will use in this analysis, the elliptic genus for ${\cal N}=(0,2)$ theories and the $\S^2 \times \T^2$ partition function for ${\cal N}=1$ theories.

\

\subsection{Elliptic genus}

Let us start by reviewing the computation of the elliptic genus of $2d$ theories with ${\cal N}=(0,2)$ supersymmetry.  This was computed in the RR sector in \cite{Benini:2013nda,Benini:2013xpa} and in the NSNS sector in \cite{Gadde:2013dda}; in this paper we will use the former convention.  We work on a torus with complex structure $\tau$.  Since the background is flat, all the supercharges are preserved, and we can attempt to localize the path integral to BPS configurations.  It was shown in \cite{Benini:2013nda} that these are given by flat connections for the gauge fields, with holonomies which we may take to lie in the Cartan.  For a $U(1)$ gauge field, we define a complexified fugacity, $z=e^{2 \pi i u}$ where $u=\int_1 A-\tau \int_2 A$.

The $1$-loop determinant of a chiral multiplet with unit charge in this background is,

\be \label{02chi}  Z_{chi}(z;q) = \frac{1}{\theta(z;q)}\,, \ee
where $q=e^{2 \pi i \tau}$ and $\theta(z;q) = q^{\frac{1}{12}} z^{-\frac{1}{2}} \prod_{k=0}^\infty(1-z q^k)(1-z^{-1} q^{k+1})$.  For a fermi-multiplet, we have,

\be \label{02ferm} Z_{ferm}(z;q) = \theta(z;q) \,.\ee
For dynamical gauge fields, we have find a contribution of:

\be \label{02vec}  Z_{gauge}(z;q) = \eta(q)^{2 r_G} \prod_{\alpha \in Ad(G)} \theta(z^\alpha;q)\,. \ee
Here $\eta(q) = q^{\frac{1}{24}} \prod_{k=1}^{\infty}(1-q^k)$, and the prefactor comes from the contribution of the gauge multiplet components along the Cartan.  We use the shorthand $z^\alpha = \prod_{j=1}^{r_G} {z_j}^{\alpha_j}$, where $r_G$ is the rank of the group, $z_j$ runs over a basis of the Cartan, and $\alpha_j$ are the components of the roots in this basis.  

After collecting the $1$-loop determinant factors for all the matter and gauge fields, which depend on $z_j$ as well as fugacities $\mu_a$ for background gauge fields coupled to flavor symmetries, we integrate this over a certain half-dimensional contour $C$ in the complex space of $z_j$'s,

\be {\cal I}(\mu_a) = \oint_C \frac{dz_j}{z_j} Z_{1-loop}(z_j,\mu_a)\,. \ee
We will describe the contour $C$ in a moment.  In order for the gauging to be well-defined, the gauged symmetry must not have any mixed 't Hooft anomalies with itself or the remaining flavor symmetries.  This condition translates in the genus to the condition that the $1$-loop determinant be elliptic in $z_j$, {\it i.e.}, invariant under $z_j \rightarrow q z_j$.  Thus it suffices to consider the function in a single fundamental domain.

The contour $C$ is determined by the Jeffrey-Kirwan (JK) residue prescription, which we briefly review.  Note from the $1$-loop determinant expressions above that poles in the integrand are contributed only by chiral multiplets, and correspond to choices of holonomies for which one or more chirals develop zero modes.  These singularities occur along codimension one subspaces in $z_j$ space, and when $r_G$ or more of these subspaces intersect at a point, we may find a non-trivial residue there.  Then the JK prescription tells us which of these residues we should count.  First, in the rank one case, the prescription is simply to count the residues in the fundamental domain from those chirals which have positive charge, or equivalently, the negative of the residues from those with negative charge.  These are the same since the sum of residues of poles in the fundamental domain of an elliptic function is zero

In the higher rank case the situation is somewhat more complicated.  To each chiral we can associate a charge vector, $Q_j$, such that the argument of the theta function corresponding to this chiral depends on the $z_j$ through $z^Q$.  Consider a point where $m \geq r_G$ of chirals develop a zero mode, {\it i.e.}, $m \geq r$ of the codimension one singular subspaces intersect, and let $Q_a$, $a=1,...,m$ be the corresponding charge vectors.  In the neighborhood of the singular point (which we may take to lie at $z_j=1$ for simplicity), the function we integrate looks like a sum of terms of the form (writing $z_j =e^{2 \pi i u_j}$),

\be  \label{jkterms} f(u) \frac{du_1}{Q_{a_1}(u)} \wedge \frac{du_2}{Q_{a_2}(u)}\wedge ...\wedge \frac{du_{r_G}}{Q_{a_{r_G}}(u)}\,, \ee
for some choice of $a_i \in \{1,...,m\}$, and with $f(u)$ holomorphic.  Then to define the JK residue, we must pick an auxiliary $r_G$-vector $\eta_j$, and then we count the contribution from this singular point as,

\be \label{jk} \mbox{JK-Res}_{u \rightarrow 0} \; ...=  \left\{ \begin{array}{cc} f(0)\; |\mbox{det}(Q_{a_1} Q_{a_2}...Q_{a_{r_G}})|^{-1} & \;\; \eta \in \mbox{Cone}(Q_{a_i}) \\ 0 & \mbox{else} \end{array} \right.\,, \ee
where $\mbox{Cone}(Q_{a_i})$ is the positive cone spanned by the $Q_{a_i}$.  The final answer, obtained by summing this over all singular points in the fundamental domain, is independent of the choice of $\eta$.  Eg, in the rank one case, the sign of $\eta$ tells us whether to count poles from positively or negatively charged chirals.

\

\subsection{$\S^2\times \T^2$ partition function}

Next we consider the computation of the $\S^2\times \T^2$ partition function of ${\cal N}=1$ theories in \cite{Benini:2015noa}.\footnote{This partition function was also computed in \cite{Honda:2015yha}, however there is an important difference in the prescription between the two papers, namely, the sum over fluxes we will review below is absent in the latter.  We find this sum plays a crucial role for the computations to be  consistent, and so we will use the prescription of the former paper below.}  An ${\cal N}=1$ theory with a suitable choice of $R$-symmetry can be placed on a certain SUGRA background on $\S^2 \times \T^2$, with a background $R$-symmetry gauge field with unit flux through the $\S^2$, while preserving two of the supercharges, as we saw for $\S^2 \times \R^2$ above.  After placing the theory on this background, one can use standard localization arguments to reduce the path integral to a finite dimensional integral and/or sum over BPS configurations.  The BPS configurations relevant here are flat connections for the gauge field on the $\T^2$, labeled by the holonomies around the two cycles, and a flux through the $\S^2$.  The flux and holonomies must all commute, and so can be labeled by two elements of the maximal torus and a coweight, respectively.  Eg, for a $U(1)$ gauge field they are labeled by a complex fugacity, $z=e^{2 \pi i u}$, $u=\int_a A - \tau \int_2 A$, as with the elliptic genus above, and an integer flux, $m = \frac{1}{2 \pi } \int_{\S^2} F$.

The partition function we compute has the following index interpretation,

\be
{\cal I}(a_\ell,s_\ell;y,q)=\Tr_{{\cal H}_{\S^2\times \S^1;s_\ell}} (-1)^F q^{L_0} y^{J^3} \prod_\ell {a_\ell}^{q_\ell}\,,
\ee where $L_0$ is the momentum along spatial $\S^1$, $J^3$ is the Cartan of the $SU(2)$ isometry of $\S^2$, $q_\ell$ are charges under global flavor symmetries, with complexified fugacities $a_\ell$, and ${\cal H}_{\S^2\times \S^1;s_\ell}$ is the Hilbert space on $\S^2 \times S^1$ with fluxes $s_\ell$ for gauge fields coupled to global symmetries.  Below we will set the fugacity $y$ to one for simplicity.

The $1$-loop determinant of an ${\cal N}=1$ chiral multiplet of R-charge $r$, which must be integer, coupled to the $U(1)$ background above is given by,

\be \label{s2t2chi} \bigg( \frac{1}{\theta(z;q)} \bigg)^{m+1-r}\,. \ee
The contribution of the gauge field is,

\be  \label{s2t2vec} \eta(q)^{2 r_G} \prod_{\alpha \in Ad(G)} \theta(z^\alpha;q)\,. \ee
Then the $\S^2 \times \T^2$ index is given by taking the $1$-loop determinant factor from the chirals and vectors, which depend on fugacities and fluxes for both the dynamical and background gauge multiplets, and summing over the fluxes and integrating over holonomies for the dynamical gauge fields,

\be {\cal I}(\mu_a,s_a) = \sum_{m_j \in \Lambda_{cw}} \oint_C \frac{dz_j}{z_j} Z_{1-loop}(z_j,m_j;\mu_a,s_a) \,.\ee
Here $z_j$ and $\mu_j$ are holonomies for dynamical and background gauge fields, as in the elliptic genus above, and $m_j$ and $s_a$ are fluxes.  Also, $C$ is a certain half-dimensional contour in the space of complexied fugacities $z_j$.  

To define this contour, let us first note that, for a given $m_j$, the integrand is precisely what appears in the the elliptic genus of a certain ${\cal N}=(0,2)$ gauge theory with matter content determined by the rules (here $G$ and $H$ are the gauge and flavor symmetry groups, respectively),

\be \label{rules} {\cal N}=1 \; \mbox{vector multiplet} \; \rightarrow\;  {\cal N}=(0,2)\; \mbox{vector multiplet } \ee
$$  {\cal N}=1 \; \mbox{chiral multiplet transforming with weight} \; (\rho,\omega) \; \mbox{under} \; G \times H \;\rightarrow  $$
$$ \rightarrow \; \left\{ \begin{array}{cc} m\;\; {\cal N}=(0,2) \ \mbox{chiral multiplets} &\;\;\; m>0 \\ |m| \;\; {\cal N}=(0,2) \; \mbox{fermi multiplets} & \;\;\;m<0 \\ \mbox{does not contribute} &\;\;\; m=0 \end{array} \right\}  \;\;\; \mbox{ with weight} (\rho,\omega) \; \mbox{under} \; G \times H $$
where $m = \rho(m) + \omega(s) +1-r$ is the flux felt by the chiral, with $r$ its R-charge.  This follows from comparing the $1$-loop determinants in (\ref{s2t2chi}) and (\ref{s2t2vec})  to those for the elliptic genus in (\ref{02chi}), (\ref{02ferm}), and (\ref{02vec}).  Then the contour $C$ is identical to the one appropriate for computing the elliptic genus of this ${\cal N}=(0,2)$ theory, {\it i.e.}, it is determined by the JK prescription.  Note if we set to zero all fluxes for the dynamical and background gauge fields, this is precisely the rule we found in (\ref{redrul}).

To summarize, we see that the $\S^2 \times \T^2$ partition function is identical to the infinite sum of elliptic genera of ${\cal N}=(0,2)$ field theories, whose matter content is determined by the matter content of the $4d$ theory, the choice of R-symmetry, and the magnetic flux of the gauge field. 

\

\subsection{Truncation of sum over fluxes}

However, an important simplification occurs, which is that this infinite sum truncates to a finite sum.  In favorable cases, it even truncates to a single term, which we can then interpret as the elliptic genus of a unique ${\cal N}=(0,2)$ theory.  Let us argue how this truncation occurs, first starting for simplicity with the rank one case.\footnote{In \cite{Closset:2015rna} it was also observed, in the context of the $A$-twisted $\S^2$ partition function, which is computed by a similar prescription, that for certain choices of the parameter $\eta$, many of the flux sectors do not contribute.  However, we expect that in this case, as well as the case of $\S^2 \times \S^1$, the sum does not, in general, truncate to a finite sum.  In particular, interpreting the $\S^2 \times \S^1$ partition function as computing the index of the $1d$ quantum mechanics we get by reducing on the $\S^2$ seems to be less straightforward than in the $4d$ to $2d$ reduction.}

\

\subsection*{Rank one case}

For a rank one gauge theory, the index is given explicitly by,

\be {\cal I}(\mu_a,s_a) = \sum_{m \in \mathbb{Z}} \oint_C \frac{dz}{z} \prod_\alpha \bigg( \frac{1}{\theta(z^{e_\alpha} {\mu_a}^{f_{\alpha,a}};q)} \bigg)^{e_\alpha m+f_{\alpha,a} s_a +1-r_\alpha}\,, \ee
where $e_\alpha$ and $f_{\alpha,a}$ are the gauge and flavor charges of the $\alpha$th chiral, and $r_\alpha$ is its R-charge.   Here the contour $C$ should be taken to encircle only those poles contributed by positively charged chirals, with $e_\alpha>0$, or equivalently, encircling only those from the negatively charged chirals (with the opposite orientation).  

Now, note that the $\alpha$th chiral only contributes a pole when,

$$ e_\alpha m+f_{\alpha,a} s_a -r_\alpha \geq 0 $$

\be \Rightarrow \left\{ \begin{array}{cc} \displaystyle m \geq \frac{1}{|e_\alpha|}(-f_{\alpha,a} s_a + r_\alpha),  & \;\;\; e_\alpha>0, \\
& \\
\displaystyle m \leq \frac{1}{|e_\alpha|}(f_{\alpha,a} s_a - r_\alpha),  & \;\;\; e_\alpha<0. \end{array} \right. \,.\ee
Thus when $m$ is sufficiently positive, the negatively charged chirals do not contribute any poles.  If we choose to count the poles coming from them, we get zero, and so we see the sum truncates for sufficiently large $m$.  Similarly, when $m$ is sufficiently negative, the positively charged fields do not contribute poles, and so if we count poles from them, we see the sum truncates below.  Thus the sum which was naively infinite actually simplifies to a finite sum.

We can get an even more drastic simplification if we do not turn on any flavor fluxes and make the key assumption that {\it the R-charge of all chiral multiplets is non-negative}.  In this case, we see that for $m>0$, none of the negatively charged chirals have poles, and similarly for $m<0$ and the positively charged chirals.  Thus in this case we have a contribution only from the term $m=0$.  This term looks like the elliptic genus of a $2d$ $(0,2)$ theory with a chiral multiplet for each field of R-charge $0$, no contribution from R-charge $1$ fields, and $r-1$ fermi multiplets for each field of R-charge $r>1$.

\

\subsection*{Higher rank case}

Next consider the higher rank case, and let us determine which fluxes may contribute in the sum defining the $\S^2 \times \T^2$ index.  Specifically, we are interested in finding the condition when only the contribution from the sector with all $m_j=0$ is non-zero, so let us fix some non-zero vector of fluxes, $m_j$, and see whether it contributes.  For simplicity, we do not turn on any flavor fluxes.  As described above, we compute the contribution from this term using the JK residue prescription, which depends on a choice of $r_G$-vector $\eta_j$, and a convenient choice for us will be to take $\eta_j=-m_j$.\footnote{More precisely, we must choose an $\eta_j$ which does not lie on the boundary of the positive cone of any set of $r_G$  $Q_{a}$ vectors, for then the prescription in (\ref{jk}) will not be well-defined.  If this choice of $\eta_j$ lies on such a boundary, we can simply deform it slightly to fix this, and this will not affect the argument below.}

Similar to above, the condition for the $\alpha$th chiral to contribute a pole is $e_\alpha^j m_j -r_\alpha\geq 0$, where $e_\alpha^j$, $j=1,...,r_G$, are the gauge charges and $r_\alpha$ is the R-charge.  Now suppose we find a point where $r_G$ chirals simultaneously get poles.\footnote{If more than $r_G$ chirals get poles, we should first write the integrand near the singular point as a sum of terms of the form (\ref{jkterms}), and apply the argument below to each of them.} Then we must have,

\be \label{ineq} e_\alpha^j m_j - r_\alpha \geq 0, \;\;\; \alpha=1,... ,r_G \,.\ee
From (\ref{jk}), in order for this pole to actually contribute, $\eta$ must lie in the positive cone of the $e_\alpha^j$, {\it i.e.},

\be \eta_j = -m_j = \sum_\alpha c^\alpha e_\alpha^j \,,\ee
where $c^\alpha>0$.  Now if we sum the inequality (\ref{ineq}) over $\alpha$, weighted by the positive coefficients $c^\alpha$, we find,

\be 0 \leq \sum_\alpha c^\alpha ( e_\alpha^j m_j - r_\alpha) = -\sum_j {m_j}^2- \sum_\alpha c^\alpha r_\alpha \,.\ee
We can see that, for sufficiently large $m_j$, this inequality will be violated (since the second term depends only linearly on the $m_j$), and so this pole can't contribute.  Thus the sum over $m_j$ must truncate to a finite sum.  If we assume, moreover, that {\it all the $r_\alpha$ are non-negative}, then we can see that only the term with all $m_j=0$ can possibly contribute.

\

To summarize, if we take a $4d$ theory for which we assign all fields non-negative R-charges,\footnote{
We also note that taking the R-charges to be in addition not bigger than two implies that the partition function does not depend on the fugacity coupled to the isometry of $\S^2$.}
 the $\S^2 \times \T^2$ index is identical to the elliptic genus of a certain $2d$ $(0,2)$ theory, with matter content determined by the rules (\ref{rules}).  In particular, a $4d$ duality will imply the identity of elliptic genera of the corresponding $2d$ theories, and can be taken as evidence for a $2d$ duality.  Let us now look at some examples.

\

\section{Reduction of Seiberg dualities}

Let us here illustrate how the simple criterion discussed above can be used to find evidence in favor of two dimensional dualities.

\

\subsection{$SU(N_c)$ dualities}

As a first example, take ${\cal N}=1$ $SU(N_c)$ SQCD with $N_f$ flavors $(Q_a,\tilde{Q}_a)$. We need to pick an assignment of non-negative integer R-charges for the flavors, which must obey the anomaly-free condition,

\be \sum_{a=1}^{N_f} r_a + \tilde{r}_a = 2(N_f-N_c) \,,\ee
where $r_a=R(Q_a)$ and $\tilde{r}_a=R(\tilde{Q}_a)$.  We also assume $N_f >N_c$.\footnote{For $N_f<N_c$, the theory is supersymmetry breaking.  For $N_f=N_c$, there is a quantum-deformed moduli space, and we expect the partition function is singular, as occurs in the superconformal index \cite{Spiridonov:2014cxa}.}  Let us pick $r_a>1$, $a=1,...,\ell$ for the first $\ell$ chirals, R-charge $1$ for the next $k$, and R-charge zero to the remaining $N_f-\ell-k$, and similarly with $\tilde{r}_a$, $a=1,...,\tilde{\ell}$ and $\tilde{k}$ for the anti-fundamentals.  The theory we find in $2d$ consists of $N_f-\ell-k$ chirals in the fundamental, $N_f-\tilde{\ell}-\tilde{k}$ chirals in the antifundamental, and $N_{ferm}$ fermis in the fundamental/anti-fundamental (these are equivalent for fermi multiplets), where,

$$ N_{ferm} =  \sum_{a=1}^{\ell} (r_a-1) + \sum_{a=1}^{\tilde{\ell}} (\tilde{r}_a-1) = 2 (N_f-N_c) - \ell - \tilde{\ell} -k - \tilde{k} $$

\be = N_{chi} -2N_c\,, \ee
where we used the anomaly-free condition, and $N_{chi}$ is the total number of chirals.  Note this relation between $N_{chi}$ and $N_{ferm}$ is required so that the $SU(N_c)^2$ anomaly cancels in $2d$.

Now consider the Seiberg dual of the $4d$ theory \cite{Seiberg:1994pq}.  This is an $SU(N_f-N_c)$ theory with $N_f$ fundamental flavors and ${N_f}^2$ mesons.  In general, given the R-charge assigments of the quarks in the original theory, we can read off the R-charge assigments of the dual fields as,

\be R(q_a) = 1- r_a + \beta , \;\;\; R(\tilde{q}_a) = 1-\tilde{r}_a- \beta, \;\;\; R(M_{ab}) = r_a+\tilde{r}\,, \ee
where $\beta = \frac{1}{2(N_f-N_c)} \sum_a (r_a -\tilde{r}_a)$, which we may assume without loss to be non-negative.  For the choice of R-charges above, one can see that non-negativity of the R-charges implies $\beta=0$ or $1$.\footnote{Reductions with no restriction of non-negativity were first discussed in \cite{Yujinote}.}  

Let us discuss the solutions we find in the two cases.  For $\beta=1$, non-negativity of the R-charges of the $\tilde{q}$'s requires $\tilde{k}=\tilde{\ell}=0$, and for $Q_a$, $a=1,...,\ell$, non-negativity requires $r_a=2$.  The anomaly-free condition is then $k+2\ell=2(N_f-N_c)$, which also ensures $\beta=1$.  Note $k$ must be even, and let us write $k=2n$.  Thus the $2d$ ${\cal N}=(0,2)$ theories we get on the two sides, which we conjecture to be dual, are:

$${\bf (A)}: \;  SU(N_c) \mbox{ with $N_c-n$ fund. chirals, $N_f-N_c-n$ fund. fermis, and $N_f$ anti-fund. chirals}  $$
$$ {\bf (B)}: \; SU(N_f-N_c) \mbox{ with $N_c-n$ fund. fermis, $N_f-N_c-n$ fund. chirals, and $N_f$ anti-fund. chirals, } $$
\be \label{b1dual} \mbox{ with $N_f (N_c-n)$ chiral mesons and $N_f(N_f-N_c-n)$ fermi mesons} \ee

For $\beta=0$, we see we must take $\ell=\tilde{\ell}=0$.  Then imposing the anomaly-free condition and $\beta=0$ gives $k=\tilde{k}=N_f-N_c$.  Thus we see the $2d$ ${\cal N}=(0,2)$ theories we get in this case, which we also conjecture to be dual, are:

$${\bf (A')}: \;SU(N_c) \mbox{ with $N_c$ fund. chirals  and $N_c$ anti-fund. chirals}  $$
$${\bf (B')}: \; SU(N_f-N_c) \mbox{ with $N_f-N_c$ fund. chirals and $N_f-N_c$ anti-fund. chirals, } $$
\be \label{b0dual} \mbox{ with ${N_c}^2$ chiral mesons and $(N_f-N_c)^2$ fermi mesons} \ee
Note $N_f$ does not appear on the first side of the duality.  If we take $N_f=N_c+1$, we find a duality to a WZ model:\footnote{Note that the earlier version of the paper contained a wrong discussion here.}

\be SU(N_c) \mbox{ with $N_c$ fund. chirals and $N_c$ anti-fund. chirals} \nonumber  \ee
\be  \label{b1dualspecial}  \leftrightarrow \;\; \mbox{  ${N_c}^2$ chiral mesons, 2 chiral baryons, and 1 fermi meson}  \ee
For general $N_f$, we can move the $(N_f-N_c)^2$ fermi multiplets in (\ref{b0dual}) to the other side by adding a mass term, and also add an additional 2 chirals and a fermi to each side, and we see it is just given by taking two instances of (\ref{b1dualspecial}).

In \cite{Gadde:2013lxa}, the authors studied similar theories, but with gauge group $U(N_c)$ rather than $SU(N_c)$.  The ranks of the flavor symmetry groups above correspond in their notation to $N_1=N_f,N_2=N_c-n,$ and $N_3=N_f-N_c-n$.  There they found a triality corresponding to cyclic permutations of $N_1,N_2,N_3$ (the remaining permutations are related to these by charge conjugation).  The duality above corresponds to the exchange of $N_2$ and $N_3$.  The theory which would naively be the third theory in the triality would have rank $-n$, which is an indication of SUSY breaking in the original theories for the $U(N)$ version of these theories (or, for $n=0$, that the theories are free).  For the $SU(N)$ versions this is no longer the case, and these theories have non-trivial fixed points, however it does imply that we are only able to find a duality, and not a triality.  
\

\subsubsection{Global Symmetries}

Let us understand how the reduction works in more detail by studying the global symmetries of these theories.  In the parent $4d$ theories there is an $SU(N_f) \times SU(N_f) \times U(1)_B \times U(1)_R$ symmetry under which the fields are charged as:
\begin{center}
\begin{tabular}{|c|c|c|c|c|}
\hline
 & $SU(N_f)$ & $SU(N_f)$ & $U(1)_B$ & $U(1)_R$ \\
\hline
$Q_a$ & ${\bf N_f}$ & ${\bf 1}$ & $\frac{1}{N_c}$ & $1-\frac{N_c}{N_f}$ \\
$\tilde{Q}_a$ & ${\bf 1}$ & ${\bf N_f}$ & $-\frac{1}{N_c}$ & $1-\frac{N_c}{N_f}$ \\
\hline
$q_a$ & ${\bf \bar{N_f}}$ & ${\bf 1}$ & $\frac{1}{N_f-N_c}$ & $\frac{N_c}{N_f}$ \\
$\tilde{q}_a$ & ${\bf 1}$ & ${\bf \bar{N_f}}$ & $-\frac{1}{N_f-N_c}$ & $\frac{N_f}{N_c}$ \\
$M_{ab}$ & ${\bf N_f}$ & ${\bf N_f}$ & $0$ & $\frac{N_c}{N_f}$ \\
\hline
\end{tabular}
\end{center}
Now let us see what happens when we redefine the R-symmetry as above.  Let us first consider the case $n>0$.  Then our choice of R-charges breaks the $SU(N_f)$ symmetry acting on $\tilde{Q}_a$ to $SU(N_c-n) \times SU(2n) \times SU(N_f-N_c-n) \times U(1)_1 \times U(1)_2$.  Following \cite{Gadde:2013lxa} we write:

\be N_1 \equiv N_c-n, \;\;\; N_2 \equiv N_f, \;\;\; N_3 \equiv N_f-N_c-n \ee and rename the fields.  Then we find:
$$ $$
\begin{center}
\begin{tabular}{|c|c|c|c|c|c|c|c|c|}
\hline
 & $SU(N_2)$ & $SU(N_1)$ & $SU(2n)$ & $SU(N_3)$ & $U(1)_1$ & $U(1)_2$ & $U(1)_B$ & $U(1)_R$ \\
\hline
$\Phi_a$ & ${\bf N_2}$ & ${\bf 1}$ & ${\bf 1}$ & ${\bf 1}$ & $0$ & $0$ & $\frac{1}{N_c}$ & $0$ \\
$P_a$ & ${\bf 1}$ & ${\bf N_1}$ & ${\bf 1}$ & ${\bf 1}$ & $2n$ & $0$ & $-\frac{1}{N_c}$ & $0$ \\
$\Delta_a$ & ${\bf 1}$ & ${\bf 1}$ & ${\bf 2n}$ & ${\bf 1}$ & $-(N_1)$ & $(N_3)$ & $-\frac{1}{N_c}$ & $1$ \\
$\Psi_a$ & ${\bf 1}$ & ${\bf 1}$ & ${\bf 1}$ & ${\bf N_3}$ & $0$ & $-2n$ & $-\frac{1}{N_c}$ & $2$  \\
\hline
$\hat{\Phi}_a$ & ${\bf \bar{N_2}}$ & ${\bf 1}$ & ${\bf 1}$ & ${\bf 1}$ & $0$ & $0$ & $\frac{1}{\hat{N}_c}$ & $0$ \\
$\hat{\Psi}_a$ & ${\bf 1}$ & ${\bf \bar{N_1}}$ & ${\bf 1}$ & ${\bf 1}$ & $-2n$ & $0$ & $-\frac{1}{\hat{N}_c}$ & $2$ \\
$\hat{\Delta}_a$ & ${\bf 1}$ & ${\bf 1}$ & ${\bf \bar{2n}}$ & ${\bf 1}$ & $N_1$ & $-(N_3)$ & $-\frac{1}{\hat{N}_c}$ & $1$ \\
$\hat{P}_a$ & ${\bf 1}$ & ${\bf 1}$ & ${\bf 1}$ & ${\bf \bar{N_3}}$  & $0$ & $2n$ & $-\frac{1}{\hat{N}_c}$ & $0$  \\
${M^P}_{ab}$ & ${\bf N_2}$ & ${\bf N_1}$ & ${\bf 1}$ & ${\bf 1}$ & $2n$ & $0$ & $0$ & $0$  \\
${M^\Delta}_{ab}$ & ${\bf N_2}$ & ${\bf 1}$ & ${\bf 2n}$ & ${\bf 1}$ & $-(N_1)$ & $(N_3)$ & $0$ & $1$  \\
${M^\Psi}_{ab}$ & ${\bf N_2}$ & ${\bf 1}$ & ${\bf 1}$ & ${\bf N_3}$ & $0$ & $-2n$ & $0$ & $2$  \\
\hline
\end{tabular}
\end{center}
$$ $$
where $N_c=\frac{1}{2}(N_2+N_1-N_3)$, and $\hat{N_c}= \frac{1}{2}(N_2+N_3-N_1)$

After reducing to $2d$, we can eliminate the R-charge $1$ fields which don't contribute, and correspondingly the $SU(2n)$ symmetry disappears.  We also move the fields $M^P_{ab}$ to the first side by adding a superpotential, and rename these to $\Gamma_{ab}$, and rename ${M^\Psi}_{ab}$ to $\hat{\Gamma}_{ab}$.  Finally, we renormalize the $U(1)_1$ and $U(1)_2$ symmetries so all fields have charge $\pm 1$.  We are finally left with a $2d$ ${\cal N}=(0,2)$ theory with fields charged as:
\begin{center}
\begin{tabular}{|c|c|c|c|c|c|c|}
\hline
 & $SU(N_2)$ & $SU(N_1)$ & $SU(N_3)$ & $U(1)_1$ & $U(1)_2$ & $U(1)_B$ \\
\hline
$\Phi_a$ & ${\bf N_2}$ & ${\bf 1}$ & ${\bf 1}$ & $0$ & $0$ & $\frac{1}{N_c}$ \\
$P_a$ & ${\bf 1}$ & ${\bf N_1}$ & ${\bf 1}$ & $1$ & $0$ & $-\frac{1}{N_c}$  \\
$\Psi_a$ & ${\bf 1}$ & ${\bf 1}$ &  ${\bf N_3}$ & $0$ & $-1$ & $-\frac{1}{N_c}$   \\
${\Gamma}_{ab}$ & ${\bf \bar{N_2}}$ & ${\bf \bar{N_1}}$ & ${\bf 1}$ & $-1$ & $0$ & $0$  \\
\hline
$\hat{\Phi}_a$ & ${\bf \bar{N_2}}$ & ${\bf 1}$ & ${\bf 1}$ & $0$ & $0$ & $\frac{1}{\hat{N_c}}$  \\
$\hat{\Psi}_a$ & ${\bf 1}$ & ${\bf \bar{N_1}}$ & ${\bf 1}$ & $-1$ & $0$ & $-\frac{1}{\hat{N_c}}$  \\
$\hat{P}_a$ & ${\bf 1}$ & ${\bf 1}$ & ${\bf \bar{N_3}}$  & $0$ & $1$ & $-\frac{1}{\hat{N_c}}$  \\
$\hat{\Gamma}_{ab}$ & ${\bf N_2}$ & ${\bf 1}$ & ${\bf N_3}$ & $0$ & $-1$ & $0$  \\
\hline
\end{tabular}
\end{center}
$$ $$

\begin{figure}[h]
  \centering
    \includegraphics[width=0.5\textwidth]{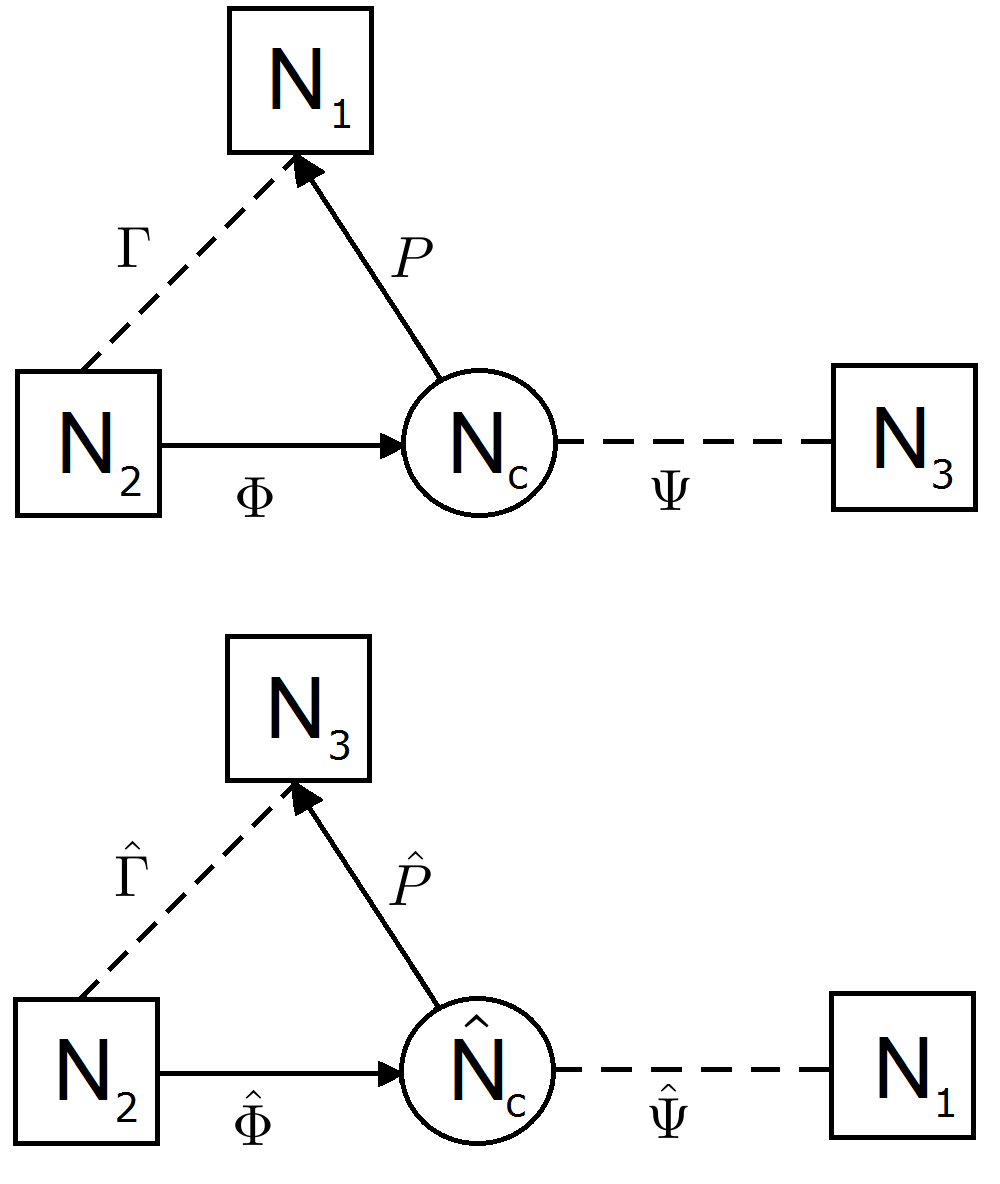}
\caption{Quivers for the dual theories.  Here the central node is an $SU(N)$ gauge group, while the outer nodes are $U(N)$ flavor groups, with $N_c=\frac{1}{2}(N_1+N_2-N_3)$ and $\hat{N}_c=\frac{1}{2}(N_3+N_2-N_1)$ .  Here we must impose $N_2 \geq N_1 + N_3$, and as a result we obtain a duality rather than a triality.}
\end{figure}

One can compute the matrix of abelian anomalies as,
\begin{center}
\begin{tabular}{|c|ccc|}
\hline
 & $U(1)_1$ & $U(1)_2$ & $U(1)_B$ \\ 
\hline
$U(1)_1$ & $\frac{1}{2} N_1(N_1-N_3-N_2)$ & $0$ & $-N_1$ \\
$U(1)_2$ & $0$ & $\frac{1}{2} N_3(N_3-N_1-N_2)$ & $-N_3$ \\
$U(1)_B$ & $-N_1$ & $-N_3$ & $2$ \\
\hline
\end{tabular}
\end{center}
We can see this is consistent with the duality, where $N_1 \leftrightarrow N_3$ and $U(1)_1 \leftrightarrow U(1)_2$.

Next consider the case $n=0$, {\it i.e.}, $N_2=N_1+N_3$.  Then the choice of R-symmetry only breaks the $SU(N_f)$ to $SU(N_1) \times SU(N_3) \times U(1)$, {\it i.e.}, there is only a single $U(1)$ factor.  The charges in $2d$ in this case are,

\begin{center}
\begin{tabular}{|c|c|c|c|c|c|}
\hline
 & $SU(N_2)$ & $SU(N_1)$ & $SU(N_3)$ & $U(1)_1$ & $U(1)_B$ \\
\hline
$\Phi_a$ & ${\bf N_2}$ & ${\bf 1}$ & ${\bf 1}$ & $0$ & $\frac{1}{N_c}$ \\
$P_a$ & ${\bf 1}$ & ${\bf N_1}$ & ${\bf 1}$ & $N_3$ & $-\frac{1}{N_c}$  \\
$\Psi_a$ & ${\bf 1}$ & ${\bf 1}$ &  ${\bf N_3}$ & $-N_1$ & $-\frac{1}{N_c}$   \\
${\Gamma}_{ab}$ & ${\bf \bar{N_2}}$ & ${\bf \bar{N_1}}$ & ${\bf 1}$ & $-N_3$ & $0$  \\
\hline
$\hat{\Phi}_a$ & ${\bf \bar{N_2}}$ & ${\bf 1}$ & ${\bf 1}$ & $0$ & $\frac{1}{\hat{N_c}}$  \\
$\hat{\Psi}_a$ & ${\bf 1}$ & ${\bf \bar{N_1}}$ & ${\bf 1}$ & $-N_3$ & $-\frac{1}{\hat{N_c}}$  \\
$\hat{P}_a$ & ${\bf 1}$ & ${\bf 1}$ & ${\bf \bar{N_3}}$  & $N_1$ & $-\frac{1}{\hat{N_c}}$  \\
$\hat{\Gamma}_{ab}$ & ${\bf N_2}$ & ${\bf 1}$ & ${\bf N_3}$ & $-N_1$ & $0$  \\
\hline
\end{tabular}
\end{center}

\

In fact naively each of these theories has an additional axial $U(1)_A$ symmetry in $2d$ under which all the quarks have the same charge. This symmetry is anomalous in the parent $4d$ theory. We thus would not expect the $U(1)_A$ to be the symmetry of the dimensionally reduced theory. A possible way for this to happen is for the dimensionally reduced theory to have a superpotential explicitly breaking $U(1)_A$ symmetry.
An analogous effect happens when reducing four dimensional theories on $\S^1$, where in general a monopole superpotential is generated in the theory in three dimensions which explicitly breaks the symmetries which were anomalous in four dimensions \cite{Aharony:2013dha,Aharony:2013kma}.  A similar phenomenon may occur here, possibly involving a local defect operator in two dimensions, such as the dimensional reduction of the codimension two Gukov-Witten operator in $4d$ \cite{Gukov:2006jk}.  Although we do not pursue this issue here, we think it would be an interesting topic for future investigation.

We have checked in many examples that the elliptic genera, refined by fugacities for all the symmetries, agree between the two theories.  Note this serves as a check not only of our proposed $2d$ duality, but also of the $4d$ Seiberg duality, for which these computations can be interpreted as the $\S^2 \times \T^2$ index with a certain choice of $U(1)_R$ symmetry. 

\

\subsection{$U(N)$ version}

Let us comment on what happens in the reduction of the $U(N)$ version of this duality.  In $4d$ we can obtain a $U(N)$ version of Seiberg duality by gauging the $U(1)_B$ flavor symmetry on both sides.  However, there is a $U(1)_B\;U(1)_B\;U(1)_R$ anomaly, so the theory has no non-anomalous R-symmetry, and so we will not be able to compactify the theory on $\R^2 \times \S^2$.  To fix this, it is convenient to introduce additional fields charged under the $U(1)_B$.  Namely, we find the following theories are dual, and have non-anomalous $U(1)_R$,

\begin{center}
\begin{tabular}{|c|c|c|c|c|}
\hline
 & $SU(N_f)$ & $SU(N_f)$ & $U(1)_B$ & $U(1)_R$ \\
\hline
$Q_a$ & ${\bf N_f}$ & ${\bf 1}$ & $\frac{1}{N_c}$ & $1-\frac{N_c}{N_f}$ \\
$\tilde{Q}_a$ & ${\bf 1}$ & ${\bf N_f}$ & $-\frac{1}{N_c}$ & $1-\frac{N_c}{N_f}$ \\
$\Omega^{\pm}$ & ${\bf 1}$ & ${\bf 1}$ & $\pm 1$ & $2$ \\
\hline
$q_a$ & ${\bf \bar{N_f}}$ & ${\bf 1}$ & $\frac{1}{N_f-N_c}$ & $\frac{N_c}{N_f}$ \\
$\tilde{q}_a$ & ${\bf 1}$ & ${\bf \bar{N_f}}$ & $-\frac{1}{N_f-N_c}$ & $\frac{N_f}{N_c}$ \\
$M_{ab}$ & ${\bf N_f}$ & ${\bf N_f}$ & $0$ & $\frac{N_c}{N_f}$ \\
$\Omega^{\pm}$ & ${\bf 1}$ & ${\bf 1}$ & $\pm 1$ & $2$ \\
\hline
\end{tabular}
\end{center}
$$ $$

If we restrict to all R-charges non-negative, we are led to the same analysis as above, and in particular we only find theories which are free or SUSY breaking.  However, suppose we impose this restriction only on the fundamentals, but do not impose any restriction on the R-charges of the anti-fundamentals.  Then we claim we can only get contributions to the $\S^2 \times \T^2$ index from fluxes $\{ m_j\}$ with all $m_j \geq 0$.  Namely, suppose that, say, $m_1<0$.  Then $\eta_1>0$, and so the only way an intersection of singular subspaces can contribute is if it involves a pole from the $1$st component of a fundamental chiral.  But given our restriction on R-charges such poles do not appear, and so there is no such contribution.

In fact we can do better, and truncate further to $m_j=0$.  Namely, in the $\S^2 \times \T^2$ index of a theory with a $U(1)$ gauge factor, we can introduce an FI parameter $\zeta$ in the action, and in the index this enters as a factor $w^m$ weighing the sum over fluxes $m$, where $w=e^{2 \pi i \zeta}$.  Using $\zeta$ we can isolate a term with fixed $m$, and argue that these must map across the duality.  In the present case this allows us to restrict to $\sum_j m_j=0$, and then from the last paragraph we see only the $m_j=0$ sector contributes here.

With this extra freedom, one can in fact obtain the full $U(N)$ triality of \cite{Gadde:2013lxa}, at the level of the elliptic genus, starting from equality of $\S^2 \times \T^2$ indices in $4d$.  Namely, in \cite{Honda:2015yha,Yujinote}, an $R$-charge assignment of the above form was picked and it was argued  that one recovers the triality. 

\

\subsection{$Sp(2N_c)$ Seiberg duality}

Next consider the $Sp$ version of Seiberg duality \cite{Intriligator:1995ne}, which relates the $Sp(2N_c)$ theory with $2N_f$ fundamental chirals, and the $Sp(2(N_f-N_c-2))$ theory with $2N_f$ fundamental chirals and $N_f(2N_f-1)$ mesons.  

We must choose an assignment of integer R-charges, $r_a$, to the fundamental chirals.  The anomaly-free condition is,

$$ \sum_a r_a = 2(N_f-N_c-1)\,. $$  
Under duality, the R-charges of the dual quarks are given as,

$$ \hat{r}_a = 1-r_a\,. $$
Thus the only way to guarantee non-negative R-charges on both sides is to take $r_a=0$ or $1$.  Then the anomaly free condition forces us to take $2(N_f-N_c-1)$ of the chirals to have R-charge $1$, and the remaining $2(N_c+1)$ to have R-charge zero, leaving a $2d$ theory with $2(N_c+1)$ chiral flavors.  On the dual side, we find a theory with $2(N_f-N_c-1)$ fundamental chirals, $(N_c+1)(2N_c+1)$ chiral mesons, and $(N_f-N_c-1)(2(N_f-N_c-1)-1)$ fermi mesons.  As in the duality of (\ref{b0dual}), this gives rise to two decoupled dual pairs and, similarly to that case, we may isolate one of them by considering $N_f=N_c+2$.  Then this leads to a 2d duality,\footnote{We are grateful to Matteo Sacchi for pointing out an error in this duality appearing in the original version of this paper. Note that for $N_c=1$ and $N_f=2$ here the duality is the same as $N_c=2$, $N_f=3$, and $n=1$ of \eqref{b1dual} and also as \eqref{b1dualspecial} for $N_c=2$. This case was studied in detail in \cite{Dedushenko:2017osi}.}

\be \label{spdual} Sp(2N_c) \mbox{ with $2(N_c+1)$ fundamental chirals}  \;\;\; \leftrightarrow \;\;\; \mbox{  $(N_c+1)(2N_c+1)$ chiral mesons, $M_{ab}$ ,} \nonumber \ee
\be \qquad\qquad\qquad\qquad\qquad\qquad \qquad \mbox{$1$ fermi meson, $\Lambda$} \ee
Thus we also find a dual WZ model in the 2d reduction of this 4d duality, but do not find dualities between two gauge theories.

\

\section{Reductions of ${\cal N}=2$ models}

Next we consider four dimensional theories with ${\cal N}=2$ supersymmetry.  Recall that a Lagrangian for such a theory can be constructed from hypermultiplets, which in ${\cal N}=1$ language consist of a pair of chiral multiplets, $q_a$ and $\tilde{q}_a$, as well as ${\cal N}=2$ vector multiplets, consisting of an ${\cal N}=1$ vector multiplet and an adjoint chiral multiplet $\Phi$.  These theories classically have an $SU(2)_R \times U(1)_r$ R-symmetry, as well as a $U(1)_{F_a}$ flavor symmetry acting on each hypermultiplet.  These act on the fields and supercharges as\footnote{Here we write the charges of the bottom component of the corresponding multiplet.},

\begin{center}
\begin{tabular}{|c|c|c|c|}
\hline
 & $U(1)_R \subset SU(2)_R$ & $U(1)_r$ & $U(1)_{F_a}$ \\
\hline
$q_b$ & $1$ & $0$ & $\delta_{ab}$ \\
$\tilde{q}_b$ & $1$ & $0$ & $-\delta_{ab}$ \\
$\Phi$ & $0$ & $2$ & $0$ \\
\hline
${\cal Q}_\alpha^\pm$ & $\pm 1$ & $1$ & 0 \\
${\cal \tilde{Q}}_{\dot \alpha}^\pm$ & $\pm 1$ & $-1$ & 0 \\
\hline
\end{tabular}
\end{center}
$$ $$
Here the $U(1)_r$ symmetry will be anomalous unless the theory is conformal.

In order to compactify the theory on $\R^2 \times \S^2$, we must pick an ${\cal N}=1$ R-symmetry.  There are several choices, leading to different theories in $2d$ with various types of supersymmetry.

\

\noindent {\bf Coulomb reduction}

First consider taking the ${\cal N}=1$ R-symmetry to be $U(1)_R$, the Cartan of the $SU(2)_R$ symmetry, which is always non-anomalous.  This gives the hypermultiplets fields $q,\tilde{q}$ an R-charge of $1$ and the adjoint chiral $\Phi$ R-charge zero.  The supercharges which are preserved are ${\cal Q}_\alpha^+$ and ${\cal \tilde{Q}}_{\dot \alpha}^+$, and the resulting theory has ${\cal N}=(2,2)$ supersymmetry.  We see the hypermultiplets do not contribute in $2d$, while an ${\cal N}=2$ vector multiplet contributes a $(0,2)$ adjoint chiral and vector multiplet, {\it i.e.}, an ${\cal N}=(2,2)$ vector multiplet.   The theory we obtain is, at least for Lagrangian theories in $4d$, a pure ${\cal N}=(2,2)$ gauge theory with the same gauge group as in $4d$.

\

\noindent {\bf ``Flavored'' reductions}

Starting from this choice, we can further mix the R-symmetry with some ${\cal N}=2$ flavor symmetry without changing the supersymmetry that is preserved.  Take a free hypermultiplet, for example.  If we mix the R-symmetry with the $U(1)_F$ symmetry with one sign, we find that $q$ gets R-charge $0$ and contributes a $(0,2)$ chiral multiplet, and $\tilde{q}$ gets R-charge $2$ and contributes a $(0,2)$ fermi multiplet, and so (after replacing the fermi multiplet by its conjugate) these contribute an ${\cal N}=(2,2)$ chiral multiplet in the representation of $q$.  If we take the other sign choice, we find a chiral in the representation of $\tilde{q}$.  In a general Lagrangian theory, we can make such a choice for each hypermultiplet in the theory, {\it i.e.}, whether it contributes a chiral multiplet in the representation of $q$, in that of $\tilde{q}$, or does not contribute at all.  A generic choice will break some of the non-abelian flavor symmetry.  All these choices give theories with ${\cal N}=(2,2)$ SUSY.

Suppose the four dimensional theory we start with is conformal.   Then the symmetry $\frac{1}{2}(R-r)$ is a non-anomalous flavor symmetry, and acts on $q,\tilde{q},\Phi$ as $\frac{1}{2},\frac{1}{2},$ and $-1$, respectively.  This becomes the the $U(1)_V$ R-symmetry of the ${\cal N}=(2,2)$ algebra, and acts on each chiral multiplet with charge $\frac{1}{2}$.
In the non-conformal case, we expect a superpotential is generated dynamically in the compactification which breaks this $U(1)_V$ symmetry, similar to the $N_2=N_1+N_3$ case of the $SU(N)$ Seiberg duality described above.

\

\noindent {\bf Higgs reduction}

 Another choice, only possible in the conformal case, is to take the R-symmetry to be $U(1)_r$.  This preserves the supercharges ${Q_\alpha}^\pm$, {\it i.e.}, $4$ supercharges of the same chirality, and so we obtain a theory with ${\cal N}=(0,4)$ supersymmetry.  The hypermultiplets fields $q,\tilde{q}$ get R-charge $0$ and the adjoint chiral $\Phi$ R-charge $2$.  Thus the hyper contributes a $(0,4)$ hypermultiplet in $2d$ and the ${\cal N}=2$ vector multiplet contributes a $(0,4)$ vector multiplet.  Note there is no further mixing we can do with flavor symmetries here, as it would result in fields of negative R-charge.  As in the ${\cal N}=(2,2)$ case, the flavor symmetry $\frac{1}{2}(R-r)$ gives rise to a subgroup of the extended R-symmetry group of $(0,4)$.  In this reduction we obtain the theories considered in \cite{Putrov:2015jpa}, which were observed to be ${\cal N}=(0,4)$ sigma models with target space (a certain bundle over) the Higgs branch of the $4d$ theory.  In the case of the $E_6$ theory, the ${\cal N}=1$ Lagrangian discussed in \cite{Gadde:2015xta} gives rise upon reduction to an ${\cal N}=(0,2)$ Lagrangian which flows to the ${\cal N}=(0,4)$ reduction of the $E_6$ theory, and this was checked to have the expected $E_6$ symmetry in \cite{Gadde:2015xta}.

\

\noindent {\bf Schur-like reductions}

Finally, again in the conformal case, suppose we take the R-symmetry as $\frac{1}{2}(r+R)$.  Here only the supercharges ${\cal Q}_\alpha^+$ survive, and so the theory has only ${\cal N}=(0,2)$ supersymmetry.  This choice assigns the adjoint chiral R-charge $1$, and so it doesn't contribute in $2d$, and so the ${\cal N}=2$ vector multiplet contributes a $(0,2)$ vector multiplet.  However, it assigns the hypers a fractional R-charge $\frac{1}{2}$.  To make their R-charge integer, we must further mix with the flavor symmetry acting on them with coefficient $\pm \frac{1}{2}$.  For one sign $q$ and $\tilde{q}$ get R-charges $0$ and $1$, and contribute a $(0,2)$ chiral in the representation of $q$, and for the other they contribute a chiral in the representation of $\tilde{q}$.   In other words, the matter content is the same as a corresponding ``flavored reduction'', but with all ${\cal N}=(2,2)$ multiplets replaced by ${\cal N}=(0,2)$ multiplets.  The symmetry $\frac{1}{2}(R-r)$ is a flavor symmetry which we call $U(1)_y$. 
With this choice and taking $y=q$, the elliptic genus of this theory has exactly the same integrand appearing in the Schur limit of the superconformal index.  However we emphasize the contour of integration here is given by the JK prescription (and is different for the different ways of mixing with the flavor symmetries), while that for the Schur index is the unit circle.

\

\subsection{ Example: $SU(N)$ $N_f=2N$ $S$-duality}

As an example, consider the $S$-duality acting on $SU(N)$ with $2N$ flavors.  This is a self-duality, but acts in a non-trivial way on the $SU(N)_a \times SU(N)_b \times U(1)_c \times U(1)_d \subset U(2N)$ flavor symmetry of the theory, exchanging $U(1)_c$ and $U(1)_d$.  Let us study the partition functions of this theories and its various two dimensional reductions.

The $\S^2 \times \T^2$ index of this theory is given by (writing only the $m_j=0$ term, as we always pick R-charges which truncate to this term),

\be \label{sun2nind} &&{\cal I}(a_j,b_j,c,d;y) =\\&& \oint \prod_j \frac{d z_j}{z_j} \bigg|_{\prod_j z_j =1 } \prod_{j=1}^{N} \prod_{\alpha=1}^N \bigg(\frac{1}{\theta(y^{1/2} z_j {a_\alpha}^{-1} c)} \bigg)^{1-r_\alpha}\bigg(\frac{1}{\theta(y^{1/2}(z_j {a_\alpha}^{-1} c)^{-1})} \bigg)^{1-\tilde{r}_\alpha}  \times \nonumber\\
&& \times \prod_{j=1}^{N} \prod_{\beta=1}^N \bigg(\frac{1}{\theta(y^{1/2} {z_j}^{-1} b_\beta d)} \bigg)^{1-r_\beta'} \bigg(\frac{1}{\theta(y^{1/2}({z_j}^{-1} b_\beta d)^{-1})} \bigg)^{1-\tilde{r}_\beta'}  \prod_{i\neq j} \frac{\theta(z_i /z_j)}{\theta(y^{-1}z_i/z_j)^{1-r_{\Phi}}}\,.\nonumber
 \ee
Then the statement of the four dimensional duality is that this is equal to the same expression with $c \leftrightarrow d$, and R-charges mapped appropriately.

Let us consider the various choices of R-symmetry described above.  In each case we will reinterpret (\ref{sun2nind}) as the elliptic genus of the two dimensional theory we obtain by reduction.  Then the four dimensional duality will imply an identity among two such elliptic genera, which we will interpret as evidence for a two dimensional duality.

If we take the Coulomb reduction, where the R-symmetry is taken to be the Cartan of the $SU(2)_R$, the dependence on flavor symmetries drops out and the identity becomes trivial.  A more interesting choice is the ``flavored'' reduction,  where we mix the $SU(2)_R$ Cartan with $U(1)_c$ and $U(1)_d$ symmetries, with the same sign.  Then we obtain,

\be
&&{\cal I}^{2d,A}(a_j,b_j,c,d;y) =\\&& \oint \prod_j \frac{d z_j}{z_j} \bigg|_{\prod_j z_j =1 } \prod_{j=1}^{N} \prod_{\alpha=1}^N \frac{\theta(y^{-1/2} z_j {a_\alpha}^{-1} c)}{\theta(y^{1/2} z_j {a_\alpha}^{-1} c)} \prod_{j=1}^{N} \prod_{\beta=1}^N \frac{\theta(y^{-1/2} {z_j}^{-1} b_\beta d )}{\theta(y^{1/2} {z_j}^{-1} b_\beta d)}  \prod_{i\neq j} \frac{\theta(z_i /z_j)}{\theta(y^{-1}z_i/z_j)}.\nonumber
\ee
This is the genus of an ${\cal N}=(2,2)$ $SU(N)$ theory with $N$ fundamental and $N$ antifundamental chirals.  The $4d$ duality implies the genus is symmetric under $c \leftrightarrow d$, or, in more standard notation, that the $U(1)_B$ baryon symmetry maps to itself with a change of sign, with all other flavor symmetries fixed.  This gives evidence for a ${\cal N}=(2,2)$ duality, which was noticed also in \cite{Putrov:2015jpa}.  Note that, since the theories have ${\cal N}=(2,2)$ supersymmetry, we can also compute their $\S^2$ partition functions as a function of twisted masses for the flavor symmetries, which comprises an independent check of this duality.  We have computed these for some low rank cases and found it has the expected symmetry under flipping the sign of the $U(1)_B$ twisted mass.

Alternatively, we can mix the R-symmetry with $U(1)_c$ and $U(1)_d$ with opposite signs.  Then we find,

\be&&{\cal I}^{2d,B}(a_j,b_j,c,d;y) =\\&& \oint \prod_j \frac{d z_j}{z_j} \bigg|_{\prod_j z_j =1 } \prod_{j=1}^{N} \prod_{\alpha=1}^N \frac{\theta(y^{-1/2} z_j {a_\alpha}^{-1} c)}{\theta(y^{1/2} z_j {a_\alpha}^{-1} c)} \prod_{j=1}^{N} \prod_{\beta=1}^N \frac{\theta(y^{-1/2} z_j {b_\beta}^{-1} d^{-1} )}{\theta(y^{1/2} z_j {b_\beta}^{-1} d^{-1})}  \prod_{i\neq j} \frac{\theta(z_i /z_j)}{\theta(y^{-1}z_i/z_j)}\,.\nonumber
\ee
This is the genus of a $(2,2)$ $SU(N)$ theory with $2N$ fundamental chirals.  The $4d$ duality implies this is symmetric under replacing $\alpha_a \rightarrow {\alpha_a}^{-1}, \beta_a \rightarrow {\beta_a}^{-1}$, and $c \leftrightarrow d$. This is one of the dualities discussed in~\cite{Hori:2006dk}.  One could also consider choices of R-symmetry which break some of the $SU(N) \times SU(N)$ flavor symmetry, but we will not discuss them here.

Next let us discuss the Schur-like reduction where we admix the diagonal combination of the $U(1)_c$ and $U(1)_d$ to the $\frac12(r+R)$ R-symmetry. The partition function becomes,

\be&&{\cal I}^{2d,C}(a_j,b_j,c,d;y) =\\&& \oint \prod_j \frac{d z_j}{z_j} \bigg|_{\prod_j z_j =1 } \prod_{j=1}^{N} \prod_{\alpha=1}^N \frac1{\theta(y^{1/2} z_j {a_\alpha}^{-1} c)} \prod_{j=1}^{N} \prod_{\beta=1}^N \frac1{\theta(y^{1/2} z_j^{-1} {b_\beta} d)}  \prod_{i\neq j} \theta(z_i /z_j)\,.\nonumber
\ee This expression is symmetric under exchanging $c$ and $d$, which corresponds to a $(0,2)$ duality of a $SU(N)$ theory with $N$ fundamental and $N$ anti-fundamental chirals.

Finally, we take the Higgs limit,

\be&&{\cal I}^{2d,D}(a_j,b_j,c,d;y) =\\&& \oint \prod_j \frac{d z_j}{z_j} \bigg|_{\prod_j z_j =1 } \prod_{j=1}^{N} \prod_{\alpha=1}^N \frac1{\theta(y^{1/2} (z_j {a_\alpha}^{-1} c)^{\pm1})} \prod_{j=1}^{N} \prod_{\beta=1}^N \frac1{\theta(y^{1/2} (z_j^{-1} {b_\beta} d)^{\pm1})}  \prod_{i\neq j} \theta(z_i /z_j)\theta(y^{-1} z_i /z_j)\,.\nonumber
\ee This is symmetric under exchanging $c$ and $d$.  This gives one of the $(0,4)$ dualities considered in \cite{Putrov:2015jpa}.  Since the two dimensional theory we obtain in the Higgs limit does not depend on a choice of flavor symmetry, it gives in some sense the most natural analogue of the ${\cal N}=2$ class ${\cal S}$ theories in two dimensions.

\

\section*{Acknowledgments}

We would like to thank O.~Aharony, C. ~Beem, M.~ Sacchi, N.~Seiberg, and Y.~Tachikawa for useful comments and discussions.  AG is supported by Raymond and Beverly Sackler Foundation Fund and the
NSF grant PHY-1314311. SSR was partially supported by Research in Theoretical High Energy Physics grant DOE-SC00010008. BW is supported by DOE Grant de-sc0009988 and the Roger Dashen Membership.

\bibliography{paptor_revised}

\bibliographystyle{JHEP}

\end{document}